# A REVIEW ON THE STRATEGIC USE OF IT APPLICATIONS IN ACHIEVING AND SUSTAINING COMPETITIVE ADVANTAGE

Kasra Madadipouya

[1]Department of Computing and Science, Asia Pacific University of Technology & Innovation

## ABSTRACT

*Information Technology applications are growing significantly fast for the past decade. Many Business organizations with emphasize on IT applications are trying to gain competitive advantages. Emrical studies also demonstrate that IT increases competitive advantage when it acts with human resources. In this paper we analyzed and review how various IT applications can be utilized to to provide a company with strategic advantages over the global marketplace competitive forces.*

## 1. INTRODUCTION

Nowadays, the major role of information systems in business encompasses the use of information technology to develop products, services, and capabilities that provide a company with strategic advantages over the global marketplace competitive forces. Different strategic ISs have been created in order to be able to support competitive position and gaining business advantages for companies. Strategic information systems can be interpreted as connected things or parts forming a complex whole, in particular that can help a company to gain a competitive advantage, to meet the strategic companies' objectives.

As authors noted in [1], IT/IS applications could be quite effective and efficient in order to establish, convert, and share knowledge management (KM) systems which effect the philosophy of the organization management and the way that members of organization are managed. Furthermore, they provide that such applications provide the basic foundation to change and improve the management and decision making system.

In addition to that, another reason which IT is utilized in various businesses are due to reduce costs, be able to produce in high qunatity with cutting down the costs as well as improve the quality of products or services [33].

Nevertheless, it is imperative to ascertain the measures that give an extra competitive advantage to certain businesses over the others[28]. Such measures can be noted by considering the different ways IT has been used in business operation by certain companies which finally resulted in gaining a competitive advantage.

In this paper we review some previous companies that used IT as a mean to gain superiority over other opponents or preserve their competitive position, but before this a brief description about the alignment between IT and business must be explained, to understand why business and IT students must have at least a superficial understanding about IT and business respectively.

The rest of the paper is organized as follows, in section2 an explanation about alignment between IT and business is given in section3 a concise literature review will be given, and finally in conclusions and future improvements are discussed.

## 2. ALIGNMENT BETWEEN BUSINESS AND INFORMATION TECHNOLOGY

Yearly, many companies are failed to utilize IT applications or projects in a way to gain competitive advantages among their competitors due to very weak connections within various IT sector and business department. The root cause in some organization is because of their top to bottom approach which they dedicate resources. Hence, in most of the cases IT budget is fixed and so unrealistic since the actual technological cost and needs of the business have been overlooked. In addition to that, another reason that causes weak connection between business unit and IT department might be due to miscommunication among the units manager which they mostly do not share organizational goals. For instance, in organizations that CTI have lack of business knowledge and poor attitude cause the opportunities will be missed and hence, failure in the companies. Lastly, lack of coordination among various units within a company might resulted in failure of using IT application properly.

## 3. LITERATURE REVIEW

In this part of the study different techniques that can be used to gain competitive advantage are described. Also a wide range of case studies are explained to some extent to show cases that used IT to achieve superiority over their rivals.

In [18], authors noted that the only approach which a company could gain competitive advantage over its competitors is to cut down the costs of product or expenditure and introducing different and innovative products and services which fascinated end consumers. Additionally, developing alliances, locking-in customers and supplier, increasing hurdles to entry to the business and using the investment properly in IT section would increase the competitive advantage effectively. However, to reach to competitive and strategic advantages in business, companies should be able to analyze various situations to be able to shape and align the company's environment [27, 29].

It is good to mention that IT applications could be beneficial for companies strategies in order to gain competitive advantage [33]. This strategic role of information systems embroils the use of information technology to develop products, services, and capabilities, enabling a company to gain advantages over the competitive forces it faces in a global marketplace [23]. IT can be utilized effectively in different approaches to bring a competitive advantage as the same system or applications can be used for strategic advantage in one organization.

### 3.1 Utilizing IT to be innovative

Information Technology can be used to uniquely and effectively improve products or services. More information availability possible could facilitate companies to build strategic IS/IT application to strength their business planning and strategies and support corporate innovation [1] as well as [27] state that Information Communication Technology give knowledge to organizations as they can generate meaningful customer data from it. [2] further note that customer data is collected every time a customer interacts with a system which is analyzed and collectively harvest information for organizations to develop business intelligence systems or applications. They further describe technology as a platform that consolidates customer needs and corporate missions.

For instance, [5] note that Southland Transportation Ltd., ( a Calgary-based school bus operator) were looking for a way to enhance their customer service system via communicating interruptions to bus schedule based on multiple factors like, availability of drivers, weather conditions, traffic heaviness in for safety purposes as well as avoid kids to wait in the very cold condition weather. The main intention of a such innovative busing system was to inform parents in advance regarding any possible delays that might occur in the bus schedule.

As a result the company built a communication system which notify parent regarding any delays in bus schedule beforehand, therefore, kids must not wait in the freezing temperature. The company developed a mobile application for BlackBerry platform with a custom-build mobile application for bus drivers to be able to report quickly and easily about the road conditions and any other hurdle which might cause delay in the schedule to parents and Southland Transportation.

With this better and unique service, they gained a competitive advantage over competitors as most parents subscribed to their services further enlarging the market share. This also outlines potential for business growth emerging out the same application compared to large firms, additional Business from existing customers, reduction in call volume at Dispatch Office and the ability to deliver unparalleled customer service.

[30] note that Google is also another company that has aligned IT innovations to its business operations. They further note that Google's continuous innovations are always at an incremental pace whilst importantly, providing value to the end user. Google has achieved this by adopting latest IT technologies like AJAX and integrated them into Google maps. Ajax integration provides a capability to Google API's map to be able to zoom and other great features without refreshing the page which as a result made Google Map the only competitor in online map.

**3.2. Using IT application to lower costs strategically**

As [4] mentions, utilizing IT application effectively can reduce businesses processes cost as well as suppliers or customers.For instance, [9] have noted that Automation System for Architectural Practices (ASAP) and Multi-phase Integrated Automation System (MITOS) are being used in the construction companies in Turkey to gain and sustain a competitive advantage. These systems are used for quick access of correct and up-to-date information, data storage, sharing of information, ease of communication, reducing costs and less use of paper. Hence, changes in technical side resulted in quality improvement, new revenues and adding value as well as less resource consumption for the company. In [32], authors mention that IT applications are able to be utilized to regulate inbound logistics with the benefits of totally elimination of storage costs. For instance, MITOS system has integrated with material supplier database with the purpose of sharing updated information easier and more frequent with material suppliers [9]. This works as a "Just In Time System" enabling the construction companies have their required materials supplied just before they completely run out of stock. This effective and efficient system drastically reduces inventory associated costs, giving them a competitive advantage.

Premier Designs Inc, a clothing manufacturing company based in USA is another Company that used and still uses IT/IS applications for its competitive advantage [17]. The Company has used IT innovations to cut costs. The company developed a mobile application, called "Premier Designs Jeweler" in which consumers and users able to design their own clothes and hence, the information is automatically updated in the company IT systems. By so doing, they cut on labor costs on the designer teams and increase operational efficiency since ideas are generated from consumers.

### 3.3. Utilizing IT to strategically promote growth

[22] suggest that information systems and IT/IS applications need to be aligned to the business strategy for them to be strategic and address business needs for the future. Positive relationships between an IT/IS and accepted financial measures of performance have to be identified, properly planned and made use of for a simple application to be strategically used [31]. Web 2.0 is can be strategically used to promote growth. This can be achieved by using social networking sites which are information systems that are currently being aligned to businesses to promote business growth, [7,8]. Twitter, for instance, as noted by [26], in an article "demystifying social media" states that the primary functions of social media are to monitor, respond, amplify, and lead consumer behavior and links them to the passage consumers take on when making purchasing decisions.

However, numerous organizations have drawn the above stated relationship and fully take advantage of social media's exclusive ability to engage with customers. For instance, PepsiCo has utilized Twitter intensively to get its customers insights through its DEWmocracy promotions resulted in of creation of new types of Mountain Dew brand [26]. They further claim that since 2008, the company's revenues have grown as it has sold more than 36 million cases the Mountain Dew Brand.

[7] also note that web 2.0 has been strategically used by Panasonic (an electronics company) to generate product ideas and answer to online feedback of its customer to demonstrate how valuable and important customer feedback can be for the company. This gives Panasonic a competitive advantage as design inefficiencies are significantly reduced. [7], further note that web 2.0 has given tacit knowledge and collaboration to online marketing companies therefore re-engineering their business processes and changing their scope too. They also claim that Procter & Gamble (an American multinational consumer goods company) uses web 2.0 to host design contests online to tap into the unique source of knowledge of their consumers. By so doing, their market cap doubled in the same time its innovation rate doubled giving them a huge competitive advantage.

### 3.4. Using IT applications for strategic differentiation

It is argued by [4] that competitive advantage can be attained by providing the utmost level of rewards through a differentiation. In [16] author mention the importance of innovation and doing things differently from competitors in business to create uniqueness with the purpose of creating unique value to customers. He further states that for an application to be unique and strategic in current organizations, it has to be customized as well as being manageable over a long term.

For example, [18] state that Federal Express (FedEx) differentiated itself by using IT as a channel of delivery and integrates innovative information technologies to perform and provide services to its customers. The company has utilized wireless technology like RFID and Bluetooth in order to be able to track goods status in their tracking system. The company has also distinguished as the first company which innovated package tracking system for customers which given a powerful competitive edge to the company.This migrating away from a purely document-driven to a database management system significantly increase data flexibility [35].
Fedex also differentiated itself by innovating downloadable windows software "Fedex desktop" which allows customers to instigate automated shipping requests from their PC.

This placing of the custom FedEx software clients' machines is a non-monetary strategic benefit for the company because it only automates the employees duties and makes it more likely for clients to use FedEx in the future because of the ease of the operation. It can be said that it's a means of locking in customers as well. This sort of integration requires the backend database mechanism in order to pull all the necessary information needed by customers to provide a better service [3].

Author suggests in [14] that integration with supply chain in FedEx system additionally increased customer switching costs as well as competitor barriers to enter. The author mention that products such as "FedEx Insight" has offered the key element of a successful supply chain in which it is beneficial in their shipping activities, hence, it is resulted in better planning, distribution of goods and resources, managing inventories and good returns in more efficient manner and also increasing customer satisfaction. These benefits attained by companies and users of such tracking systems are efficiency, productivity, and increased communication between customers and the business. All these resulted in customer lock-in.

[32] note that technological development is one of the value chain activities a business performs to gain profits. It can therefore be deduced that Fedex has used this component to automate its processes and gain a strategic advantage.

In [21] author notes that Samsung company made innovation via utilizing automated proven methodologies by leveraging on Invention Machine software with its capability of semantic knowledge and integrated problem analysis of CD design. By utilizing such a solution, the company has gained 33% more reliability as well as reducing manufacturing bonding points from 38 to 26 percent. In addition to that, the company productivity improved by 38% as the number of adjustment reduced from 13 to 8 points.

### 3.5. Utilizing IT to develop alliances and cut costs

The strategic use of IT has enabled organizations to create alliances and gain a competitive advantage over competitors via linking there IS systems to the internet and extranets that support strategic business relationships with suppliers and customers. However, management should be aware of evolution methods that are already in place for alignment to be feasible. They further note that for alliances to be feasible, there must be compatible IT systems between alliancing companies.

For example, Ingram (known to be world's leading wholesaler of books and other printed materials, with incredible volumes of merchandise daily) previously collaborated with diverse mail providers conditional on the destination. However, the arrangement was resulted in operational challenges because of incompatibilities among systems and lack of financial justification. Ingram opted for a mailing partner who could help the company cut postage costs and in turn render a competitive advantage for itself.

Ingram Book Group partnered with DHL International in 2004 for its massive shipping and mailing needs. This partnership made it possible for services to be tailored. This ability to customize services facilitated Ingram in achieving the enhanced tracking visibility it looked-for when dealing with international shipments [13]. The partnership enabled competitive rates thereby allowing Ingram to achieve its prime goal of significantly cutting postage cost.

In [34], author stresses that by applying GPS the operational flow, tracking and monitoring of trucks would be increased significantly since they are able to estimate the delivery time and knowing about the route and status of trucks and driver scheduler. This was of the great help in improving services as they could tell which routes to use at a particular time.

It can therefore be noted that the usage of strategic information systems gave a competitive advantage because it brought about the operational effectiveness of the company compared its competitors. Ingram dramatically cut it costs and at the same time, the partnership provided a mutual benefit for both Ingram and DHL

Intel a technology company whose mission is to use IT to deliver business value to intel strategically uses IT to increase employee productivity, facilitate growth and break geographical barriers. Intel has cut its travel costs by using video conferencing (an IT application) to facilitate collaboration between widely geographically dispersed team [20].

### 3.6. Utilizing IT to lock-in customers and create a high switching cost

Information technology is also being used by certain businesses to lock in customers and suppliers through building valuable new relationships with them. This is achieved by creating inter-organizational information systems and applications which are interlinked through electronic telecommunications networks on terminals and computers of businesses. This enabled collaboration between business customers and suppliers results in new business alliances and partnerships.

For instance, Amazon, as noted by [4], has developed mobile applications that can be used in various devices for their customers and indirectly, locked them in. The reached to this target by implementing and developing IT application system like Digital Restrictions Management (DRM) that restricts users to be able to read their e-books in other platforms. Customers always have to use Amazon's applications and services. In [3] author mentions that Amazon's the one of the most complex as well as efficient supply chain in the world.

In addition to that, users should open accounts to be able to purchase which this makes the ability to build customer relationship [4]. Through such user accounts, they share certain information with their customers such as reviews and product specifications. The customer loyalty and satisfaction they provide are means of customer lock-in. The Amazon website uses of certain information technologies, such as personalization technologies to recommend products to customers based on their previous purchases and a one-click system for fast checkout. This system enables customers to enter credit card numbers and addresses at once. With later subsequent visits, they can simply click once to make a purchase, without having to re-enter information again [6].

Another company using IT to lock-in customers is Cisco. According to [11], they are using a tailored a desktop application known as Cisco Unified Personal Communicator, for PC or Mac. The system provides an interface for instant messaging, voice messaging, video conference, call conference and corporate directory. They are using this application for partners who need to access applications in the Cisco Unified Communications suite. As these applications can only be used by Cisco partners with Cisco which is a means of locking-in customers.

IT/IS can also be used to develop a strategic information base that can render information to support the firm's competitive strategies. Information bases are a strategic resource that can be used to support strategic planning, marketing, and other strategic initiatives. This availability of information breaks traditional, time, geographic, cost and structural barriers through electronic linkages with customers, suppliers, and other business entities.

26

With IT/IS, communication is much faster and information is readily available to remote locations upon request. This reduces the cost of business operations including labor costs, the number of distribution centers. For instance, companies are able to reduce the cost of distribution, inventory, production or communications.

An organization can also use IT applications strategically differentiate itself through Agility which is how fast an organization responds to business environments. Agile companies are highly dependent on Internet technologies to integrate and manage their business processes whilst providing the information-processing power to treat their many customers as individuals.

Computer Aided Design systems could change operational agility by improving the accuracy, speed of accomplishing in a cost effective and innovative manners. For Instance, according to [15], by late 2008 Zara's sales edged ahead of most of its competitors such as Gap, making it the world's largest fashion retailer. By utilizing CAD, the average time for Zara concept from idea to launch in store was 15 days.

[8,24]suggested that Organizational agility, or the ability to execute innovations and competitive moves with speed, surprise, and competitive disruption can lead a firm to a competitive advantage. They further note how fast Zara's agility was reflected in 2005, when Madonna played a set of concerts in Spain, and by the final show, teenage girls were surprised when they saw the Zara had already copied Madonna's outfit designed and made it ready to be sold. With ready to use templates and design samples that Zara has implemented in its software, Zara is now twelve times faster than its competitors and can have its product from design to manufacture and distribution in less than two weeks.

### 3.7. Utilizing IT applications to improve lead-time

Lead-time is the time taken (delay) from initiation to execution of various organizational processes. Lead time reduction could be beneficial in the terms of lower safety stocks, more forecasting accuracy and less stock out level by getting smaller orders sizes that improves inventory mechanism and more cost effective as a result [25].

IT/IS applications have been and are being used to reduce lead-time to gain a strategic competitive advantage. For instance, according to [12] Mercedes-Benz Egypt is the only Benz vehicle manufacturing company in the region that started its operation in 1998. Their traditional ordering process was very time consuming and inefficient, it took about seven days for a single order to be completed. In 2010, the company integrated Microsoft Dynamics GP enterprise resource management software and Microsoft Dynamic CRM solution to their existing applications. This resulted in cutting down the order fulfillment time by 85 per cent.

After implementation, the dealers that needed new cars can now just place their orders through highly optimized CRM applications that are now connected to the company's database. All the relevant information is already printed on the application such as price and car identification numbers and all dealers need to fill is the car model and quantity. The finance department is also connected and has access to Microsoft Dynamics GP, allowing them to directly monitor the orders and compare them with the company's credit line.

Finally, through the automated application, invoices are generated, printed and now the cars are ready to be shipped. The Lead-Time was decreased significantly since the integration of the applications. Various benefits that Mercedes-Benz gained through this integration are as reducing ordering time processing, better locating and monitoring in stock vehicles, less paperwork, faster shipments and cost reduction. This has positioned the company in greater advantage against its rivals.

### 3.8. Utilizing IT application to increase production

Critical dependence on an internally oriented system for production, sales or a service may be regarded as strategic. Just In Time systems are currently depended on by many companies which have gained competitive advantage against their competitors. JIT concept was first introduced by Taiichi Ohno to improve Toyota's competitiveness in car business industry and soon other Japanese industry implemented this method to their exciting policy [10]. Moreover, in [10] authors define JIT system as "an inventory control philosophy whose goal is to maintain just enough material in just the right place at just the right time to make just the right amount of product". Benefits such as faster response, inventory reduction as well as improvement in operations and efficiency are the results of implementing JIT in Chinese automobile industry. [19] have illustrated the importance of implementing a JIT system, how beneficial it can be with reference to the Automotive Industry in China.

They state that, Chinese manufacturers have realized the importance of JIT and manage to implement the JIT in order to maintain their operations effectiveness. An example given by the authors in this case is a company that was established and had implemented JIT in 2002. One year later they managed to position themselves in top ten motor companies in terms of its sales in the country with 50,000 units. By 2004 they managed to triple the production and by 2006 they increased their units to 300,000 positioning the company in top five motor companies in the Chinese automotive industry market. The special application that was designed to easily connect the company to its suppliers and retail stores was the key factor. They no longer needed warehouses to store their parts and materials and therefore their costs were decreased dramatically. The company also gained advantage of the power of suppliers through the system in which, suppliers could also monitor the sales and immediately was notified of any changes in respect to retailer's inventory and therefore they had to reduce their prices to convince the company to purchase from them.

Dell is another company that has harnessed and aligned IT applications to its business operations (Arthur P., February 1994). The Company uses a JIT system implemented in 1994. However, the systems remain strategic because of continuous improvements being done to the system and also that it has aligned the application to its business by customizing the system.

## 4. CONCLUSION

It is vital to note that no industry can currently survive without the use of any IT/IS system. Due to the continuous developments and changes in the IT sector, it can also be deduced that most of the strategic moves adopted by various companies no longer become strategic with time. This happens when competitors have adopted similar strategies and nothing, therefore differentiates the mode of business operations of various companies. Some strategic management and continuous improvements to the strategic systems must be adopted for long term success.

IT/IS systems are a success contributing factor mostly if customers are involved and the system is used strategically and at the right time. Customers provide information which facilitates an appropriate analysis of the processes, timescales and resources which act as the base of any continuous improvement initiative. However, IT applications need to be aligned to business strategies and factors such as performance, features, reliability, durability, serviceability, conformance to business environment, perceived quality must be taken into consideration.

If well aligned to the business, IT can support a variety of strategic objectives, including redesign of innovative applications and business processes. It also links organizations with their business partners and facilitates sharing information. Costs can dramatically be reduced as well and acquiring of competitive intelligence can be fully supported.

IT applications have enabled various companies that have used them strategically to adapt their business model to the business environment changes as well as providing other intangible benefits such as customer loyalty, improved decision making process, business process re-engineering as well as changing the business scope..